\documentclass[final,3p,times,twocolumn]{elsarticle}
\usepackage{amsmath}
\usepackage{amsfonts} 
\usepackage{graphicx}
\usepackage{epsfig}
\usepackage{subfig}
\begin{document}
\begin{frontmatter}

\title{Identifying meaningful clusters in malware data}
%
%
\author[mymainaddress]{Renato Cordeiro de Amorim\corref{mycorrespondingauthor}}
\ead{r.amorim@essex.ac.uk}
\author[mymainaddress]{Carlos David Lopez Ruiz}
\ead{cl18417@essex.ac.uk}
\address[mymainaddress]{School of Computer Science and Electronic Engineering, University of Essex, Wivenhoe Park 
CO4 3SQ, UK.}

\cortext[mycorrespondingauthor]{Corresponding author}

\begin{abstract}
Finding meaningful clusters in drive-by-download malware data is a particularly difficult task. Malware data tends to contain overlapping clusters with wide variations of cardinality. This happens because there can be considerable similarity between malware samples (some are even said to belong to the same family), and these tend to appear in bursts. Clustering algorithms are usually applied to normalised data sets. However, the process of normalisation aims at setting features with different range values to have a similar contribution to the clustering. It does not favour more meaningful features over those that are less meaningful, an effect one should perhaps expect of the data pre-processing stage.

In this paper we introduce a method to deal precisely with the problem above. This is an iterative data pre-processing method capable of aiding to increase the separation between clusters. It does so by calculating the within-cluster degree of relevance of each feature, and then it uses these as a data rescaling factor. By repeating this until convergence our malware data was separated in clear clusters, leading to a higher average silhouette width. 
\end{abstract}

\end{frontmatter}
%
%
%
%
\section{Introduction}

The term malware is used to describe malicious software, which has been designed with the specific purpose of exploiting vulnerabilities in computer systems. In general, any given malware aims at compromising such systems. Malware can be further divided into non-exclusive categories such as trojan, virus, adware, worms, etc. Malware development may have had an innocent start, but there are now multiple examples suggesting this is a multi-million business sometimes associated to organised crime \cite{NHS.BBC,Ransomware.BBC,Ransomware2.BBC,Ransomware3.BBC}. Malware has also been used in acts of sabotage, and may even have political motivations \cite{PoliticalMalware.BBC}. 

Here, we are interested in clustering malware data. That is, identifying $k$ homogeneous groups (ie. clusters) of malware in a given data set --- without the need for labelled samples for an algorithm to learn from. Once one is able to say that a new malware sample should be assigned to a cluster containing homogeneous malware samples (for a longer discussion of what a cluster is, see \cite{hennig2015true}) it becomes easier to create defence mechanisms. Clustering algorithms (for complete reviews, see \cite{mirkin2012clustering,aggarwal2014data} and references therein) can be usually divided in two groups: partitional and hierarchical. The latter includes algorithms able to produce a clustering as well as information regarding the relationships that exist between clusters (this information can be represented with the help of a dendrogram). Partitional clustering identify $k$ disjoint clusters in a given data set, so that each object (a malware sample, in our case) in the data set is assigned to a single cluster. Although hierarchical algorithms produce information one would usually associate with families (ie. a tree), it comes with a computational cost. In 2018 alone, 246,002,762 new malware variants were found \cite{istr2019}. Hence, we see partitional clustering as a more realistic approach for the real-world.

There are various examples of clustering algorithms applied to malware data in the literature (for instance, \cite{faridi2018performance,asquith2016extremely,amorim_komisarczuk} and references therein). However, these apply classical normalisation to the data sets (for details, see Section \ref{Sec:Norm}). This type of normalisation (eg. $z$-score, range normalisation, unit length, etc.) aims at putting all features used to describe an object at the same level. It does not favour more meaningful features over those that are less meaningful. There is considerable similarity between malware samples (at times some are even said to belong to the same family). Hence, it is reasonable to expect that there will be a considerable amount of overlap between clusters. Also, malware samples have a tendency to be released in bursts with a skewed distribution \cite{song2016learning}. This scenario makes clustering particularly difficult.

In this paper we introduce a novel method to deal with the problem described above. Our method is capable of increasing the separation between clusters during the data pre-processing stage. It does so by calculating the within-cluster degree of relevance of each feature in a given data set, and using this as a rescaling factor. By iterating this process our method increases the quality of malware clusterings, as measured by the average silhouette index \cite{rousseeuw1987silhouettes}. We apply our method to drive-by-download malware data. This referrers to malware delivered to client systems that browse resources on the web, usually via http. This is a very timely issue given that in 2019 Symantec has found one in ten unique resource locators (urls) to be malicious \cite{istr2019}.

The remainder of this paper is organised as follows. Section \ref{Sec:RelatedWork} discusses the clustering algorithms that are directly relevant to our research, as well as a method to measure how good a clustering is. Section \ref{Sec:Norm} briefly explains the classical normalisation algorithms used in the pre-processing stage. Section \ref{Sec:NewMethod} explains our method, together with its mathematical motivation. Sections \ref{Sec:Data} and \ref{Sec:Results} explain our process of data gathering, our methodology, and the results we have obtained. Finally, Section \ref{Sec:Conclusion} presents our conclusions, and indications of future work.

\section{Related work}
\label{Sec:RelatedWork}
Given our objective, the work we discuss in this section relates to clustering algorithms one could use to cluster malware samples. Partitional clustering algorithms aim at identifying a set $S=\{S_1, S_2, ..., S_k\}$ so that each $S_l \in S$ contains homogeneous objects, and 
$\forall(S_l,S_j \in S), \text{ } l\neq j \iff S_l \cap S_j = \emptyset$. $K$-means \cite{macqueen1967some} is arguably the most popular such algorithm \cite{jain2010data,steinley2006k}. Given a data set $\mathcal{X}$ containing $n$ objects, each described over $d$ features, $k$-means minimises the within-cluster distance
\begin{equation}
W(S,Z) = \sum_{l=1}^k \sum_{x_i \in S_l} \sum_{v=1}^d (x_{iv} - z_{lv})^2,
\label{Eq:kmeans}
\end{equation}
where $z_l \in Z$ is the centroid of cluster $S_l \in S$, that is, the $d$-dimensional point with the lowest sum of distances to all objects in $S_l$. We defined $z_l$ as a point to make it clear that it may or may not belong to $\mathcal{X}$. The $k$-means algorithm iteratively minimises (\ref{Eq:kmeans}) with three simple steps:
\begin{enumerate}
	\item Select $k$ objects from $\mathcal{X}$ uniformly at random, and use their values to initialise $z_1, z_2, ..., z_k$.
	\item Assign each $x_i \in \mathcal{X}$ to the cluster $S_l$ represented by the centroid $z_l$ that is the nearest to $x_i$.
	\item Update each $z_l \in Z$ to the centre of $S_l$. 
\end{enumerate}
The $k$-means criterion (\ref{Eq:kmeans}) applies the Euclidean squared distance. Hence, the centre of a cluster $S_l$ is the component-wise mean of its objects, that is $z_{lv} = |S_l|^{-1} \sum_{x_i \in S_l} x_{iv}$ for $v=1, 2, ..., d$.

As popular as $k$-means may be, it does have known weaknesses. The most relevant in this paper are: (i) the final clustering depends heavily on the initial set of centroids, which are usually found at random. Suboptimal initial centroids are likely to lead the algorithm to a local minima solution; (ii) $k$-means requires the number of clusters, $k$, to be known beforehand; (iii) all features are treated as if they were equally relevant, which is rather unlikely in real-world data sets.

There has been a considerable research effort to address the weaknesses above. For instance, $k$-means++ \cite{arthur2007k} selects one object $x_i \in \mathcal{X}$ uniformly at random, and then copies its values to the first centroid $z_1$. All other initial centroids are selected following a weighted probability that is proportional to the distances between objects and their nearest centroid. 

\begin{enumerate}
	\item Set $l=1$. Select an object from $\mathcal{X}$ uniformly at random and copy its values to $z_l$.
	\item Increment $l$ by one. Select an object $x_j \in \mathcal{X}$ at random, with probability $\frac{D(x_j)^2}{\sum_{x_i \in \mathcal{X}} D(x_i)^2}$ and copy its values to $z_l$, setting $Z = Z \cup \{z_l\}$.
	\item Repeat the steps above until $l=k$.
	\item Run $k$-means using each $z_l \in Z$ as an initial centroid.
\end{enumerate}
In the algorithm above $D(x_i)$ is the distance between $x_i \in \mathcal{X}$ and its nearest centroid $z_l \in Z$. Experiments show that $k$-means++ has a faster convergence to a lower criterion output (\ref{Eq:kmeans}) than the traditional $k$-means algorithm \cite{arthur2007k}. This algorithm has enjoyed considerable popularity, and it is now the default $k$-means option in MATLAB \cite{MATLAB:2019} and scikit-learn \cite{scikit-learn}.

The intelligent $k$-means algorithm ($ik$-means) \cite{mirkin2012clustering} addresses both weaknesses (i) and (ii) by identifying good initial centroids for $k$-means as well as the number of clusters in $\mathcal{X}$. This algorithm iteratively identifies anomalous clusters, afterwards the centroids of these anomalous clusters are then used as initial centroids for $k$-means. In this, $k$ is set to the number of anomalous clusters in $\mathcal{X}$.
\begin{enumerate}
	\item Set $c$ to the centre of $\mathcal{X}$, and $Z^{\prime} = \emptyset$. Identify $x_t \in \mathcal{X}$, the object that is the furthest from $c$.
	\item Apply $k$-means to the data set using $x_t$ and $c$ as initial centroids, but do not allow $c$ to move in the cluster update step. This will lead to clusters $S_t$ and $S_c$ with centroids $z_t$ and $c$, respectively.
	\item If $|S_t| \geq \theta$, set $Z^{\prime} = Z^{\prime} \cup \{z_t\}$. In any case, remove each $x_i \in S_t$ from $\mathcal{X}$. 
	\item If $|\mathcal{X}|>0$ go to Step 1. Otherwise run $k$-means by setting $k=|Z^{\prime}|$, and using each $z_l \in Z^{\prime}$ as an initial centroid.
\end{enumerate}

In the above $\theta$ is a user-defined parameter that helps to avoid small clusters in $S$, should this be of interest to the user. If the value of $k$ is known one can sort the elements of $Z^{\prime}$ by the cardinality of their initial clusters (ie. their value of $|S_t|$ in Step four), and keep only the $k$ elements of $Z^{\prime}$ with the highest cardinality (this would happen between Steps three and four). Another approach would be to select the $k$ elements of $Z^{\prime}$ in the order they were found. This way they would be the $k$ most anomalous initial centroids.

We can see that $ik$-means identifies each centroid $z_t \in Z^{\prime}$, and related cluster $S_t$ by iteratively minimising
\begin{equation}
\label{Eq:P(U,Z)_1}
P(S,Z) = \sum_{x_i \in S_t} \sum_{v=1}^d (x_{iv} - z_{tv})^2 +  \sum_{x_j \in S_c} \sum_{v=1}^d (x_{jv} - c_v)^2,
\end{equation}
where $c$ is the component-wise mean of $\mathcal{X}$.

In order to address all three weaknesses (i), (ii) and (iii), we introduced the intelligent Minkowski weighted $k$-means ($imwk$-means) \cite{de2012minkowski}. This extends $ik$-means by following the intuitive idea that a given feature $v$ may have different degrees of relevance at each cluster $S_l \in S$. We model this behaviour by introducing $w_{lv}$, the weight of feature $v$ at cluster $S_l$. The higher $w_{lv}$ is, the higher the contribution of $v$ at cluster $S_l$ is to the clustering. First, we define the weighted Minkowski distance between $x_i$ and $z_l$ as
\begin{equation}
\label{Eq:MinkDist}
d(x_i, z_l) = \sum_{v=1}^d w_{lv}^p |x_{iv}-z_{lv}|^p.
\end{equation}
The above is in fact the p$^{th}$ power of the Minkowski distance, which is analogous to the use of the Euclidean squared distance in $k$-means. This approach saves the computational effort of calculating p$^{th}$ roots, and does not change the clusterings produced by the algorithm. The $imwk$-means algorithm minimises
\begin{equation}
W(S,Z,W) = \sum_{l=1}^k \sum_{x_i \in S_l} \sum_{v=1}^d w_{lv}^p |x_{iv} - z_{lv}|^p
\label{Eq:mwk}
\end{equation}
subject to
\begin{equation}
\label{Eq:mwkConditions}
\begin{cases} 
S_l \cap S_j = \emptyset$ for $l,j = 1, 2, ..., k$ and $l\neq j;\\
w_{lv} \geq 0$ for $l=1, 2, ..., k$ and $v=1, 2, ..., d;\\ 
\sum_{v=1}^d w_{lv}=1$ for $l=1, 2, ..., k;\\
p \geq 1.
\end{cases}
\end{equation}
Leading to
\begin{equation}
\label{Eq:weights}
w_{lv} = \left( \sum_{u=1}^d \left[ \frac{D_{lv}}{D_{lu}} \right]^{1/(p-1)} \right)^{-1},
\end{equation}
where $D_{lv}$ is the dispersion of feature $v$ at cluster $S_l$, given by $\sum_{x_i \in S_l} |x_{iv} - z_{lv}|^p$. We usually add a small constant to each dispersion ($0.001$, say) to avoid a division by zero in (\ref{Eq:weights}) when $v$ perfectly discriminates $S_l$ (ie. for all $x_i \in S_l$, $x_{iv}$ has the same value). The $imwk$-means algorithm can be described as follows.
\begin{enumerate}
	\item Set $c$ to be the Minkowski centre of $\mathcal{X}$, $\mathcal{X}^{\prime}$ to be a copy of $\mathcal{X}$, and each $w_{lv} = d^{-1}$.
	\item Find the object $x_t \in \mathcal{X}^{\prime}$ that is the farthest from $c$ using (\ref{Eq:MinkDist}), and copy its values to $z_t$. 
	\item Assign each $x_i \in \mathcal{X}^{\prime}$ to either $S_t$ or $S_c$, depending on which centroid is the nearest to $x_i$ ($z_t$ or $c$) as per (\ref{Eq:MinkDist}). If this step does not change either $S_t$ or $S_c$, go to Step 6.
	\item Update $z_t$ to the Minkowski centre of its cluster $S_t$. Update each $w_{lv}$ as per (\ref{Eq:weights}). Go back to Step 3.	
	\item If $|S_t| \geq \theta$, add $z_t$ to $Z$ and $w$ to $W$. In any case, remove all objects $x_i \in S_t$ from $\mathcal{X}^{\prime}$. If $|\mathcal{X}^{\prime}|>0$ go to Step 2.
	\item Assign each $x_i \in \mathcal{X}$ to the cluster $S_l$ whose centroid $z_l$ is the nearest to $x_i$ as per (\ref{Eq:MinkDist}). If this step produces no change to to any $S_l \in S$, stop.
	\item Update each $z_l \in Z$ to the Minkowski centre of its cluster $S_l$. Update each $w_{lv}$ as per (\ref{Eq:weights}). Go back to Step 6.		
\end{enumerate}

The Minkowsi centre of a feature $v$ at cluster $S_l$ with an exponent $p$ is the value $\mu$ leading to the lowest $\gamma(p)=\sum_{x_i \in S_l} |x_{iv} - \mu|^p$. Notice $\gamma(p)$ is a convex function. Hence, one can approximate its minimum by setting $\mu = |S_l|^{-1} \sum_{x_i \in S_l} x_{iv}$, and then keep moving $\mu$ by a small number (0.0001, say) to the side that minimises $\gamma(p)$.

If one knows how many clusters a data set has, we can re-state step 5 as ``Add $z_t$ to $Z$ and $w$ to $W$. Remove all objects $x_i \in S_t$ from $\mathcal{X}^{\prime}$. If $|\mathcal{X}^{\prime}|>0$ go to Step 2''. This would require a new step between 5 and 6: ``Keep in $Z$ and $W$ only the elements related to the $k$ clusters with the highest cardinality.'', which is the approach used in the original paper (see \cite{de2012minkowski}). Of course, very much like in $ik$-means it is also possible to remove all but the first $k$ tentative centroids from $Z$.

A suitable Minkowski exponent $p$ can be found using a consensus clustering approach \cite{de2017minkowski}. This requires one to run $imwk$-means with values of $p$ from $1.1$ to $5.0$ in steps of $0.1$, leading to 40 clusterings. The chosen $p$ is that of the clustering with the highest average similarity to all other 39 clusterings, usually measured using the Adjusted Rand Index (ARI) \citep{rand1971objective}. Given two clusterings $S=\{S_1, S_2, ..., S_k\}$ and $U=\{U_1, U_2, ..., U_r\}$, the ARI is defined as
\begin{equation}
ARI(S, U) = \frac{\sum_{ij} \binom{n_{ij}}{2} - \left[\sum_i \binom{a_i}{2} \sum_j \binom{b_j}{2}\right] / \binom{n}{2} }{ \frac{1}{2} [\sum_i \binom{a_i}{2} + \sum_j \binom{b_j}{2}] - [\sum_i \binom{a_i}{2} \sum_j \binom{b_j}{2}] / \binom{n}{2} },
\end{equation}
where $n_{ij} = |S_i \cap U_j|$, $a_i = \sum_{j=1}^r |S_i \cap U_j|$, $b_j = \sum_{i=1}^k |S_i \cap U_j|$. The ARI is corrected for chance.

As well as being able to cluster a data set, one must be able to decide whether a given clustering represents the actual structure of the data set without the use of labels. Clustering validity indices (CVIs) are usually used for this purpose. There is no clear evidence in the literature showing a particular CVI to be the best in all cases, however, the average Silhouette width \cite{rousseeuw1987silhouettes} usually performs well \cite{arbelaitz2013extensive}.

For a given $x_i \in S_l$, let $a(x_i) = \frac{1}{|S_l|-1} \sum_{x_j \in S_l\setminus\{x_i\}} \sum_{v=1}^d (x_{iv}-x_{jv})^2$. That is, $a(x_i)$ is the average distance between $x_i$ and all other objects in its cluster. A low $a(x_i)$ indicates the suitability of the assignment of $x_i$ to $S_l$. Let $b(x_i) = \min\limits_{S_t \neq S_l} \frac{1}{|S_t|}\sum_{x_j \in S_t} \sum_{v=1}^d (x_{iv} - x_{jv})^2$. That is, $b(x_i)$ is the average distance between $x_i$ and the objects of its closest neighbouring cluster. A high $b(x_i)$ indicates the unsuitability of assigning $x_i$ to the closest cluster to $S_l$. The silhouette index of $x_i$ is given by

\begin{equation}
s(x_i) = \frac{b(x_i) - a(x_i)}{\text{max}\{a(x_i), b(x_i)\}}.
\end{equation}

Clearly, $-1 \leq s(x_i) \leq 1$. A $s(x_i)$ close to one indicates $x_i$ is closer to the other objects in its cluster than to objects in other clusters. We can expand this measure to deal with all $x_i \in \mathcal{X}$, that is $\frac{1}{n} \sum_{x_i \in \mathcal{X}} s(x_i)$. 

\section{Classical data normalisation}
\label{Sec:Norm}

Often, data sets contain features with different variances. Features with a higher variance will have a higher average distance than features with a lower variance. Hence, the former will have a higher contribution to the clustering than the latter. This common issue highlights the importance of data pre-processing. In this paper, we normalise our data set (for details on the data set itself see Section \ref{Sec:Data}) using
\begin{equation}
\label{Eq:Stand}
x_{iv}^{\prime} = \frac{x_{iv} - \bar{x}_v}{max(x_v) - min(x_v)},
\end{equation}
where $\bar{x}_v = \frac{1}{n} \sum_{x_i \in \mathcal{X}} x_{iv}$, the average of feature $v$ over all objects in $\mathcal{X}$. The $z$-score is also a popular choice in this scenario, it is given by
\begin{equation*}
x_{iv}^{\prime} = \frac{x_{iv} - \bar{x}_v}{\sigma_v},
\end{equation*}
where $\sigma_v$ is the standard deviation of $v$ over all objects in $\mathcal{X}$. We favoured range normalisation (\ref{Eq:Stand}) over the $z$-score because the latter is biased towards features following a unimodal distribution. This is perhaps easier to explain with an example. Let the features $v_1$ and $v_2$ be unimodal and multimodal, respectively. The standard deviation of $v_2$ will be higher than that of $v_1$, thus, the $z$-score of $v_1$ will be higher than that of $v_2$. Thus, $v_1$ will have a higher contribution to the clustering than $v_2$. However, in clustering we would be more interested in the clusters' information in $v_2$.

Another interesting characteristic of (\ref{Eq:Stand}) is that if $v$ is a binary feature, then $max(x_v) - min(x_v)=1$. Hence, the standardised value $x_{iv}^{\prime}$ is just $x_{iv} - \bar{x}_{v}$. Note that $\bar{x}_{v}$ is in fact the frequency of $v$ in the data set $\mathcal{X}$. The higher the frequency of $v$ the lower the standardised value $x_{iv}^{\prime}$, and the lower is its contribution to the clustering. This is well-aligned with intuition, a feature that is commonly present (ie. frequent) is less likely to be discriminative.

\section{Iterative cluster-dependent feature rescaling}
\label{Sec:NewMethod}

The normalisation discussed in Section \ref{Sec:Norm} sets all features of a given data set to have about the same contribution to the clustering. This can also be seen as a disadvantage because it means that features with a higher relevance are set to have the same contribution to the clustering as features with a lower relevance. Intuition indicates that features with a higher relevance to the clustering should have a higher contribution. In fact, we can go even further. A given feature $v$ may have different degrees of relevance at each cluster $S_k \in S$, and this should be taken into account during the clustering task.

We can interpret $w_{lv}$ in the distance measure used in $imwk$-means (\ref{Eq:MinkDist}) as the degree of relevance of feature $v$ at cluster $S_l$. Such assertion requires further analysis of $imwk$-means. This algorithms aims to produce a weight $w_{lv}$ for $l=1, 2, ..., k$ and $v=1, 2, ..., d$, minimising (\ref{Eq:mwk}) subject to the conditions in (\ref{Eq:mwkConditions}). Notice that the dispersion of $v$ at cluster $S_l$ is given by 
\begin{equation*}
	D_{lv} = \sum_{x_i \in S_l}  |x_{iv} - z_{lv}|^p,
\end{equation*}
allowing us to re-write (\ref{Eq:mwk}) as
\begin{equation*}
	P(U,Z, W) = \sum_{v=1}^d  \sum_{l=1}^k w_{lv}^p D_{lv}.
\end{equation*}
The Lagrangian function of the above is
\begin{equation*}
\mathcal{L}(W, \lambda)= \sum_{v=1}^d w_{lv}^p D_{lv} + \lambda\left(1-\sum_{v=1}^d w_{lv} \right).
\end{equation*}
Allowing us to equate its two partial derivatives to zero.
\begin{equation}
\label{Eq:FirstPartial}
\frac{\partial \mathcal{L}}{\partial w_{lv}} = pw_{lv}^{p-1}D_{lv} - \lambda = 0,
\end{equation}
\begin{equation}
\label{Eq:SecondPartial}
\frac{\partial \mathcal{L}}{\partial \lambda} = 1-\sum_{v=1}^d w_{lv} = 0.
\end{equation}
We can re-arrange (\ref{Eq:FirstPartial}) to
\begin{equation}
\label{Eq:FirstPartial_2}
w_{lv} = \left(\frac{\lambda}{pD_{lv}}\right)^\frac{1}{p-1},
\end{equation}
and substitute (\ref{Eq:FirstPartial_2}) into (\ref{Eq:SecondPartial})
\begin{equation*}
\sum_{v=1}^d \left( \frac{\lambda}{pD_{lv}}\right)^\frac{1}{p-1}=1.
\end{equation*}
The above leads to 
\begin{equation*}
\left(\lambda\right)^\frac{1}{p-1} = \frac{1}{\sum_{v=1}^d \left(\frac{1}{pD_{lv}}\right)^\frac{1}{p-1}}, 
\end{equation*}
and 
\begin{equation*}
w_{lv} = \left( \sum_{u=1}^d \left[ \frac{D_{lv}}{D_{lu}} \right]^{1/(p-1)} \right)^{-1}.
\end{equation*}
The weights calculated as per the above Equation minimise (\ref{Eq:mwk}) by modelling the within-cluster degree of relevance of each feature. This is quite interesting because it allows us to go a step beyond the normalisation described in Section (\ref{Sec:Norm}) by using these weights as feature rescaling factors. This is quite unusual because each feature $v=1, 2, ..., d$ will have $k$ factors, but it is fine because these are the weights minimising the clustering criteria (\ref{Eq:mwk}).

Given a set of weights calculated as per (\ref{Eq:weights}), we can re-scale a data set $\mathcal{X}=\{x_1, x_2, ..., x_n\}$ that has been normalised (see Section \ref{Sec:Norm}) using
\begin{equation}
\label{Eq:ReScale}
x^{\prime}_{iv} = w_{lv} x_{iv} \text{ for } i=1,2, ..., n, \text{ and } v=1, 2, ..., d,
\end{equation}
where $x_i \in S_l$. In other words, the rescaling factor applied to $x_{iv}$ depends on both: (i) feature $v$; (ii) the cluster $x_i$ belongs to.

The method we introduce in this paper also has another novelty. In the first step of $imwk$-means there is no data transformation that separates clusters, and each $w_{lv}$ is set to $d^{-1}$. While this seems sensible as a starting point to minimise (\ref{Eq:mwk}), it also means this starting point is suboptimal. To address this, the main part of our method iterates between generating clusterings with $imwk$-means and rescaling the data set using (\ref{Eq:ReScale}). This way, at each iteration $imwk$-means starts from a better position. Given that $\sum_{v=1}^d w_{lv}=1$ for $l=1, 2, ..., k$, each time the data set is rescaled the values of its entries are lowered. To avoid computational issues related to dealing with very small numbers, we also normalise the data set using (\ref{Eq:Stand}) between each of these main iterations.\\

\emph{Iterative cluster-dependent feature rescaling} (icdfr):
\begin{enumerate}
	\item For each value of $p$ from 1.1 to 5.0 in steps of 0.1, generate a clustering and a set of weights using $imwk$-means.
	\item Calculate the similarity between each pair of clusterings. This similarity can be calculated using the Adjusted Rand Index. Select as optimal $p$ that which is associated to the clustering with the highest average similarity to all other clusterings.
	\item Rescale the data set using the weights generated with the optimal $p$ and (\ref{Eq:ReScale}).
	\item Normalise the data set using (\ref{Eq:Stand}).
	\item Apply $imwk$-means to the new data set with the optimal $p$. This will update the weights as well as the clustering. Unless a pre-determined number of iterations has been reached (or the algorithm has converged), go to Step 3.
\end{enumerate}

In the above, steps one and two relate to a consensus approach that has been shown to find suitable values for the Minkowski exponent $p$ \cite{de2017minkowski}. Regarding the number of iterations in Step five, we experimented with 100 although the algorithm would converge much before that.

\section{Malware data}
\label{Sec:Data}

In this paper we analyse drive-by download malware. In other words, malicious code downloaded unintentionally to the user's computer. In order to gather useful data we need to release a malware sample in a safe environment, analyse the malware itself and keep track of any changes it does to such environment. Luckily, there are a number of options in terms of software we could use to accomplish this. This type of software is commonly referred as malware sandbox, and it is used to execute untrusted programs without risking the host machine (for details see \cite{gandotra2014malware,chakkaravarthy2019survey}, and references therein). Here we have chosen to use Cuckoo Sandbox 2.06 \cite{Cuckoo} mainly because it is a free open-source solution, which has been consistently used in research (see for instance \cite{barakat2014malware,vasilescu2014practical,shijo2015integrated}). However, one should note that the significance of this choice is rather low as our method does not depend on the sandbox in use (see Section \ref{Sec:NewMethod}). All we need is data describing the malware to be analysed. Hence, one can use any malware sandbox capable of fulfilling this requirement.

\begin{figure}[t]
	\centering
	\includegraphics[scale=0.4]{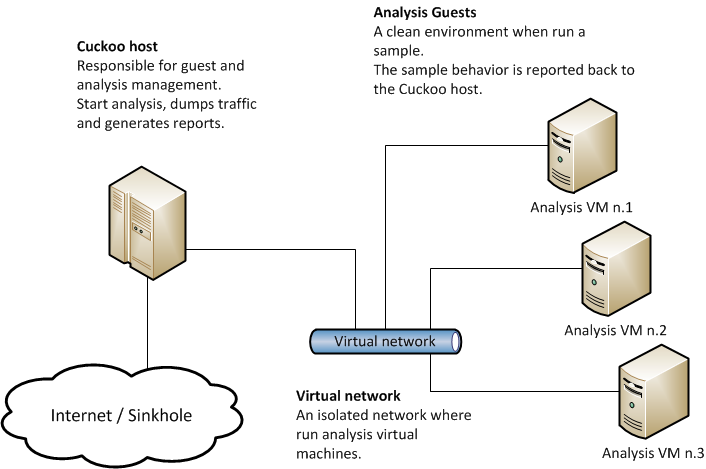}
    \caption{Cuckoo's main architecture. The host runs the management component while the guests run isolated Windows 8 environments. Each environment safely executes a malware, analyses the results, and then gets re-setted to its original (clean) state. Figure from \cite{Cuckoo}.}
    \label{Fig:Cuckoo}
\end{figure}

Cuckoo runs at host-level and manages one or more Windows 8 VM guests (see Figure \ref{Fig:Cuckoo} for a visual representation). The latter is an isolated environment allowing Cuckoo to gather behavioural data (eg. API calls made by the malware, dropped files, processes spawned, etc.). Cuckoo resets this VM to its original (ie. clean) state before each experiment with a potential malware. This particular sandbox is also able to extract information from files as part of its static analysis. For each malware Cuckoo lists a number of features related to the behaviour of the malware, as well as its static analysis. 

In terms of raw data (ie. the malware samples themselves), we acquired a total of 2,000 samples from VirusSign \cite{VirusSign}. These malware samples were gathered by VirusSign using HoneyPots, submissions, as well as trading and exchange. Each and everyone of them was confirmed by VirusSign to be malicious, using several mainstream AntiVirus software. Each malware also presented features from Cuckoo's behavioural and statical analyses.

Given a list of features obtained with Cuckoo (see Table \ref{Tab:Features}), we can transform our data into an actual data matrix. The process is quite straightforward. First we must note we have two types of features: (i) binary features, which represent the presence or absence of a particular feature at a particular malware (eg. whether or not a malware checks if the cursor is in use); (ii) numerical features, which represent the number of times a particular feature was present at a particular malware (eg. the number of times a malware sent ICMP messages). We have 2,000 samples with 67 features meaning that any given malware sample $x_i$ in our data set $\mathcal{X}$ is described over 67 features (ie. $n=2,000$ and $d=67$). 
\begin{table*}[ht] \footnotesize
\setlength{\tabcolsep}{1pt}
\begin{center}
\caption{The list of features identified by Cuckoo in the malware samples we obtained from VirusSign. In the features description a `\#' means `number of', implying the feature is numerical, all other features are binary. `PE' means Portable Executable.} 
\label{Tab:Features} 
\begin{tabular}{lllllll} 
Name&&Description&&Name&&Description\\
\cline{1-3}
\cline{5-7}
ICMP&&\# ICMP messages&&
AllocateVMem&&\# calls to NtAllocateVirtualMemory\\	
AntiDebug&&Use of debugging techniques&&
Bind&&\# calls to bind\\
CheckCursor&&Whether a cursor is in use&&
CloseSocket&&\# calls to closesocket\\
CreateFile&&\# calls to NtCreateFile&&
CreateMutant&&\# calls to NtCreateMutant\\
CryptographyReg&&Access to Cryptography registry&&
CustomLocaleReg&&Access to CustomLocale registry\\
DelayExe	&&Call to NtDelayExecution&&
DeviceIO	&&Use of DeviceIOControl\\
DroppedFiles	&&\# files dropped&&
FindFile	&&\# calls used to locate files\\
FreeVMem	&&\# calls to NtFreeVirtualMemory&&
GetSysTime&&Use of GetSystemTimeAsFileTime\\
HttpOpenReq&&\# calls to HttpOpenRequest&&
HttpSendReq&&\# calls to HttpSendRequest\\
IE&&Access to IE registry&&
MapView&&\# calls to NtMapViewOfSection\\
OpenFile	&&\# calls to NtOpenFile&&
OpenMutant&&\# calls to NtOpenMutant\\
ProcessNum&&\# processes spawned&&
ProtectVMem&&\# calls to NtProtectVirtualMemory\\
QueryFile&&\# queries for information about files&&
RegCreate&&\# calls to create registry keys\\
RegQuery&&\# queries to the registry&&
SafeBootReg&&Access to SafeBoot registry\\
Socket&&\# calls to socket&&
SortingReg&&Access to Nls/Sorting registry\\
TcpipReg	&&Access to TCP/IP registry&&
WriteFile&&\# calls to NtWriteFile\\
IReadFile&&\# calls to InternetReadFile&&
ToolSS&&Use of CreateToolHelp32Snapshot\\
Dllsloaded&&\# dlls loaded&&
SysInfoRef&&Access to SystemInformation registry\\
CryptDecodeObjectX&&\# calls to CryptDecodeObjectX&&
Fips&&Access to FIPS algorithm policy\\
CryptCreateHash&&\# calls to CryptCreateHash&&
CryptHashData&&\# calls to CryptHashData\\
DnscacheReg&&Access to DNSCache registry&&
RegModify&&\# calls to modify registry keys\\
DockingInfo&&Access to DockingState registry&&
Persistence&&Access to persistence-related registry keys\\
SCManager&&Access to service control manager&&
CryptExportKey&&\# calls to CryptExportKey\\
CryptGenKey&&\# calls to CryptGenKey&&
AppInit&&Access to DLL-loading registry keys\\
CryptAcquireContextA&&\# calls to CryptAcquireContextA&&
RemoteThread&&Creation of remote threads.\\

Files Recreated&&\# recreated files&&
DuplicateProcess&&\# call to duplicate process\\
NumSections&&\# sections in PE file&&
NumResources&&\# resources in PE file\\
NumExports&&\# exports in PE file&&
TCP&&\# tcp packets detected\\
UDP&&\# udp packets detected&&
HTTP&&\# http packets detected\\
DroppedBuffers&&\# dropped buffers&&
Yara-embedded\_ pe&&Detects embedded PE file\\
Yara-LnkHeader&&Detects lnk header&&
DnscacheReg&&DnsCache registry modified\\
Yara-embedded\_ win\_ api&&Detects embedded win api&&
Yara-shellcode&&Detects shellcode in file\\
Yara-vm\_ detect&&Use VM detection techniques&&
CheckDiskSize&&Call to CheckDiskSize function\\
Yara-embedded\_ macho&&Detects Mach-o file
\end{tabular}
\end{center}
\end{table*}
\section{Clustering results}
\label{Sec:Results}

We began by applying classical normalisation (see Section {\ref{Sec:Norm}) to the data set we constructed (for details on the data set, see Section \ref{Sec:Data}). Figure \ref{Fig:OriginalData} shows the plot of our data over its first and second principal components. Unfortunately, there is no clear evidence of a cluster structure from a Gaussian perspective (ie. clearly separable round clusters).

As popular as it may be, a clustering algorithm such as $k$-means++ will identify clusters even if there is no cluster structure in a data set. Hence, one should not just jump into applying this algorithm to the data. To illustrate this, we applied $k$-means++ to our data set 100 times. Figure\ref{Fig:kmeans} shows the clustering we obtained with the lowest criterion output by setting $k=7$ (given this is just illustrative, the actual value of $k$ matters very little). The fact this clustering is meaningless is further reinforced by an average silhouette index of $0.39$.
\begin{figure*}[t]
	\centering
    \subfloat[Normalised data set.]{
		\label{Fig:OriginalData}
    		\includegraphics[scale=0.3]{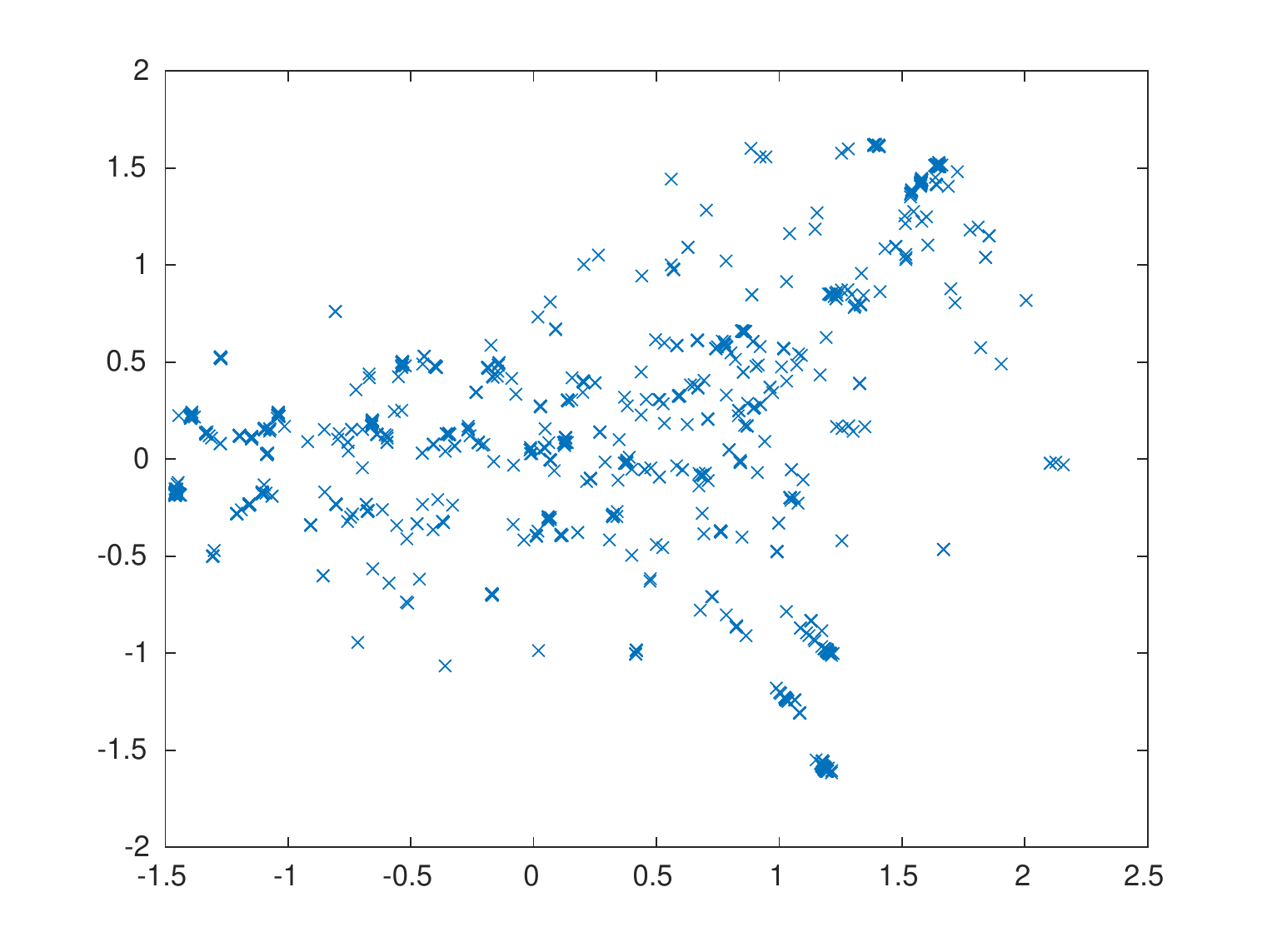} 
    		}%
    \subfloat[The $k$-means++ clustering with the lowest criterion output over 100 runs, with seven clusters. This clustering has an average silhouette index of $0.39$]{
    		\label{Fig:kmeans}
    		\includegraphics[scale=0.3]{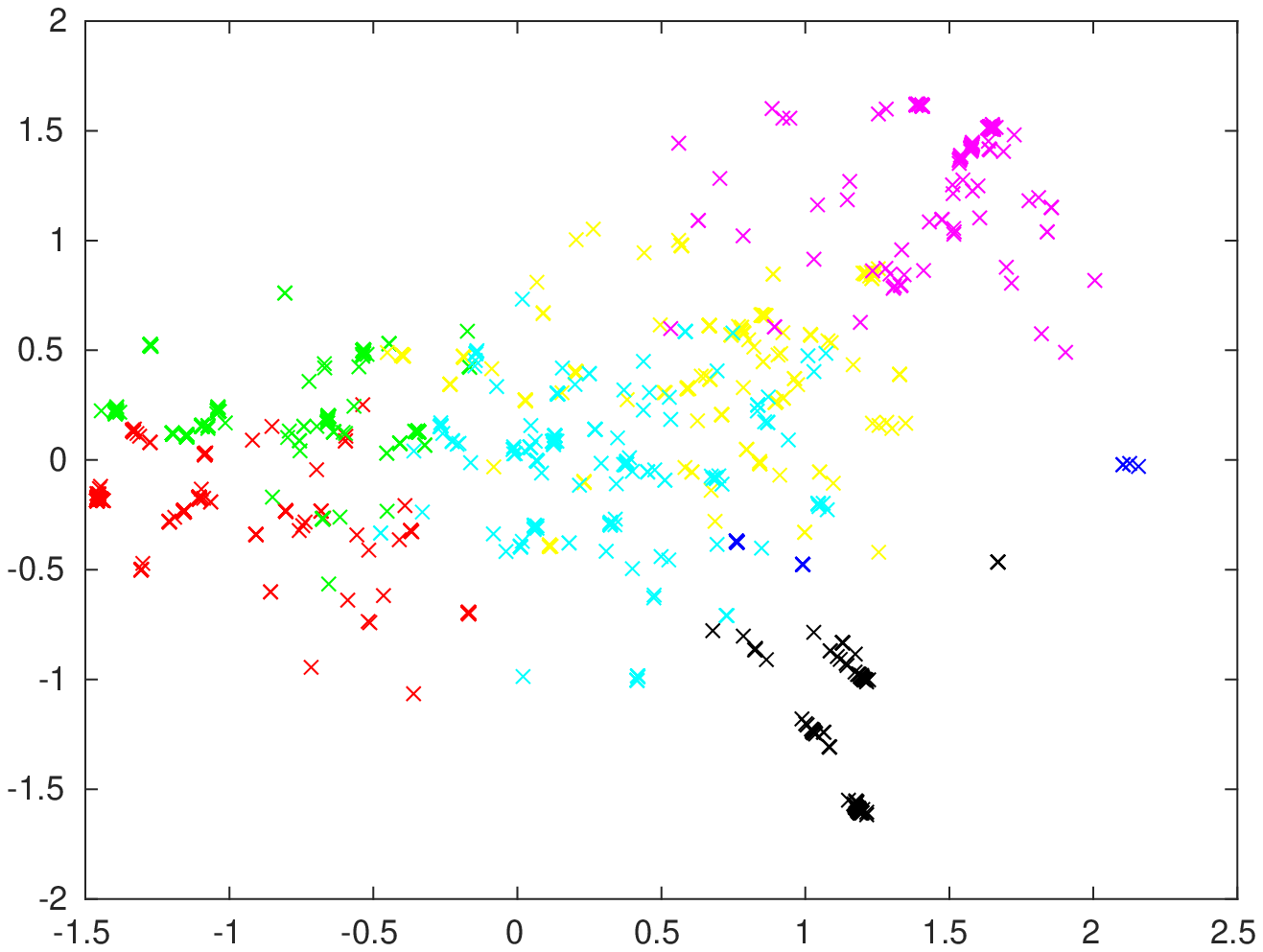} 
    		}%
    \caption{The normalised malware data set plotted over its first and second principal components.}
    \label{Fig:InitialData_std}
\end{figure*}

The results we obtained using our icdfr method are certainly more promising. In these experiments we set the $imwk$-means' thereshold $\theta=3$. We did so to avoid very small clusters in our results, which would not be of particular interest (this choice ended up leading to seven clusters, hence the number of clusters in our illustrative $k$-means++ example). The first and second steps of our method (described in Section \ref{Sec:NewMethod}) use consensus clustering in order to identify a suitable Minkowski exponent $p$ between 1.1 and 5.0 --- higher values of $p$ tend to remove the advantages of feature weighting as (\ref{Eq:weights}) will produce more uniform weights. In this experiment the optimal value was found to be $p=3.9$. Given this value of $p$ we set the number of iterations to 100 and allowed our method to follow Steps 3-5. We were happy to see the method converged in iteration 53 --- and even happier to see that the average silhouette of $imwk$-means on the data set produced by our method was 0.92. The new data set generated by our method also increased the average silhouette of $k$-means++ to 0.52.

Figure \ref{Fig:NewMethodIterations} shows our method in action. Each of its sub-figures (\textit{a} to \textit{f}) shows the plot of a $imwk$-means clustering (over its first and second principal components) on a data set generated by our method at a different (but increasing) iteration. We can clearly see that our method starts from a chaotic scenario and it quickly starts separating clusters. 
\begin{figure*}%
    \centering
    \subfloat[Iteration 1]{{\includegraphics[scale=0.3]{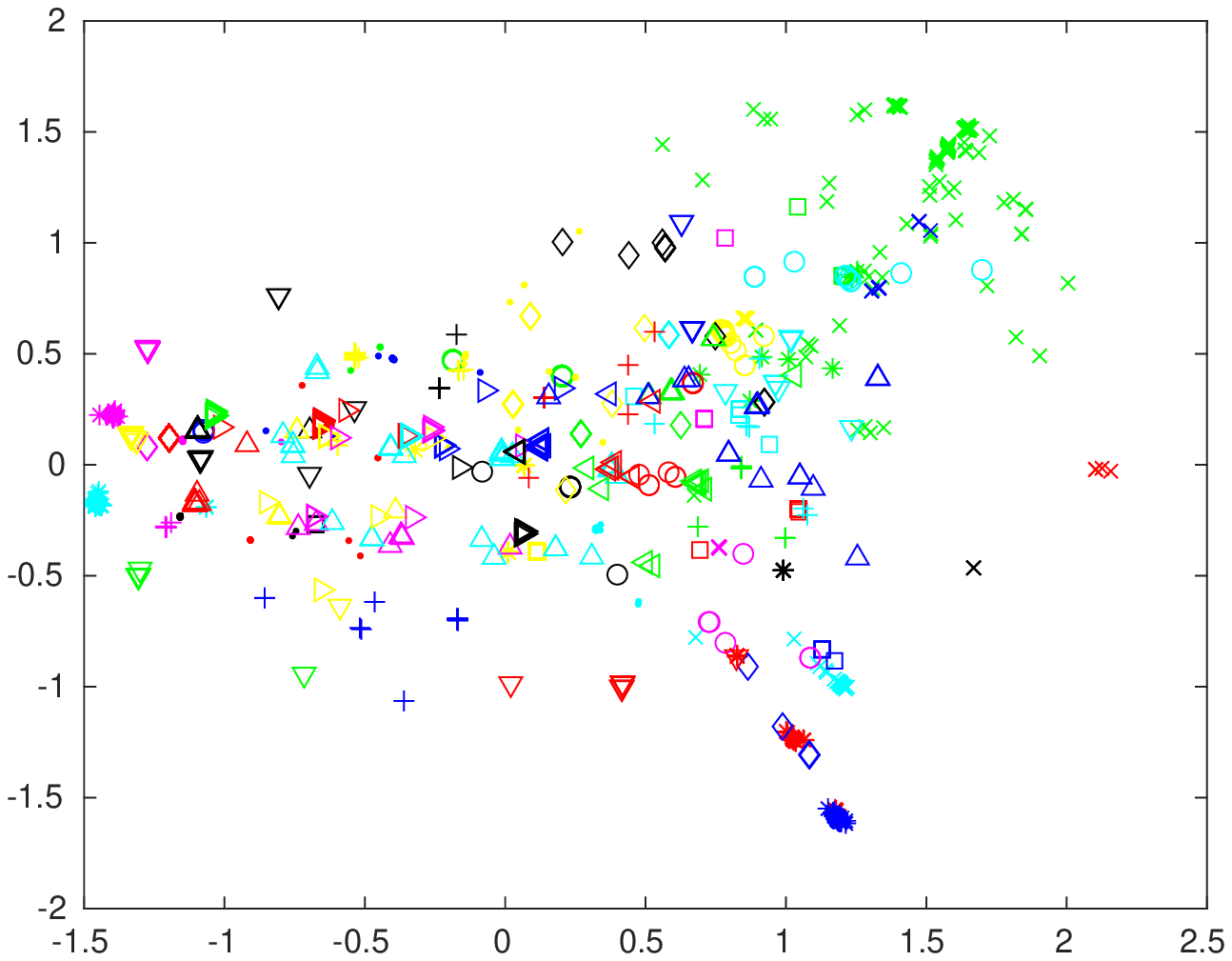} }}%
    \subfloat[Iteration 5]{{\includegraphics[scale=0.3]{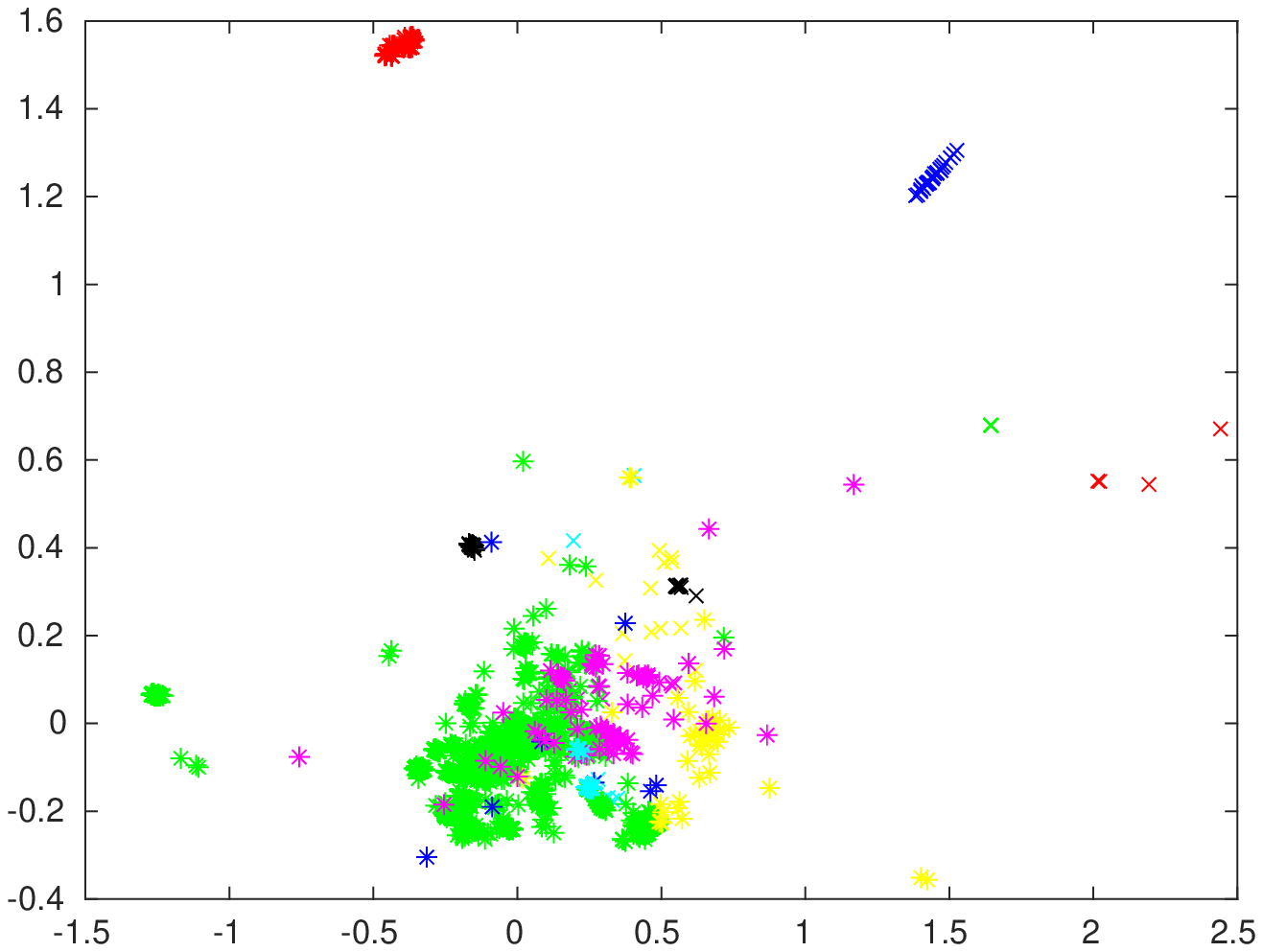} }}%
    \subfloat[Iteration 8]{{\includegraphics[scale=0.3]{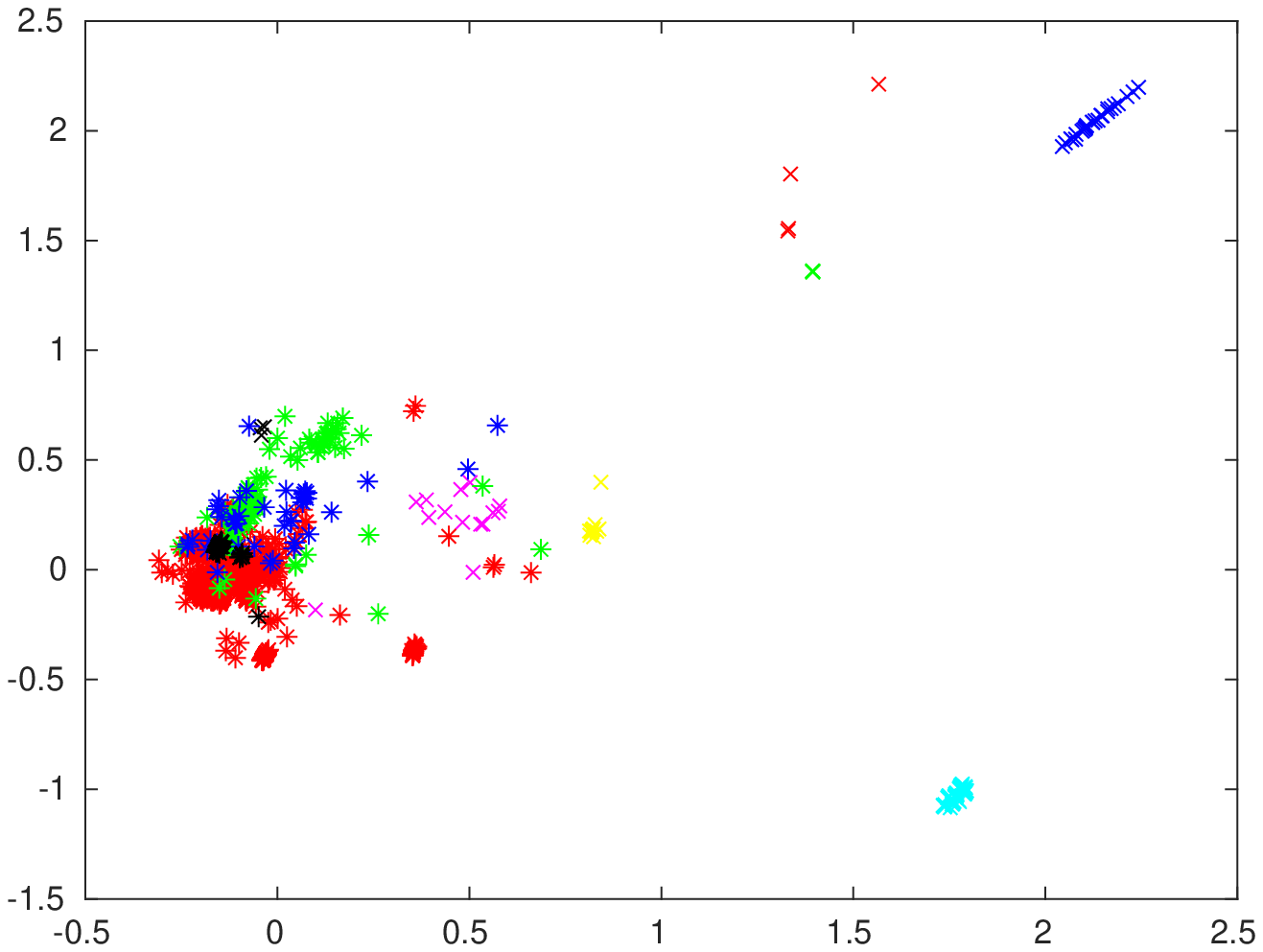} }}\\
    \subfloat[Iteration 10]{{\includegraphics[scale=0.3]{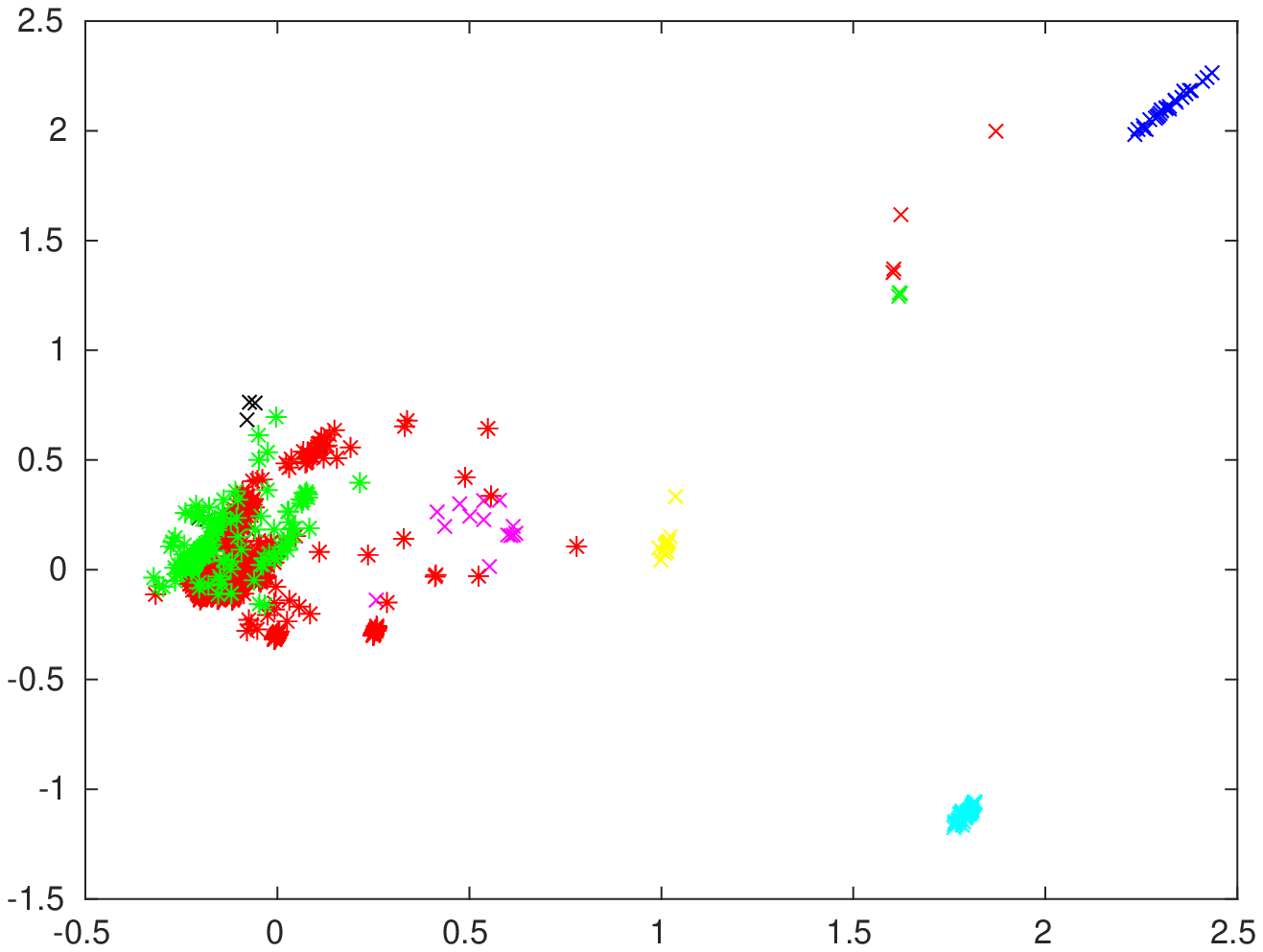} }}
    \subfloat[Iteration 29]{{\includegraphics[scale=0.3]{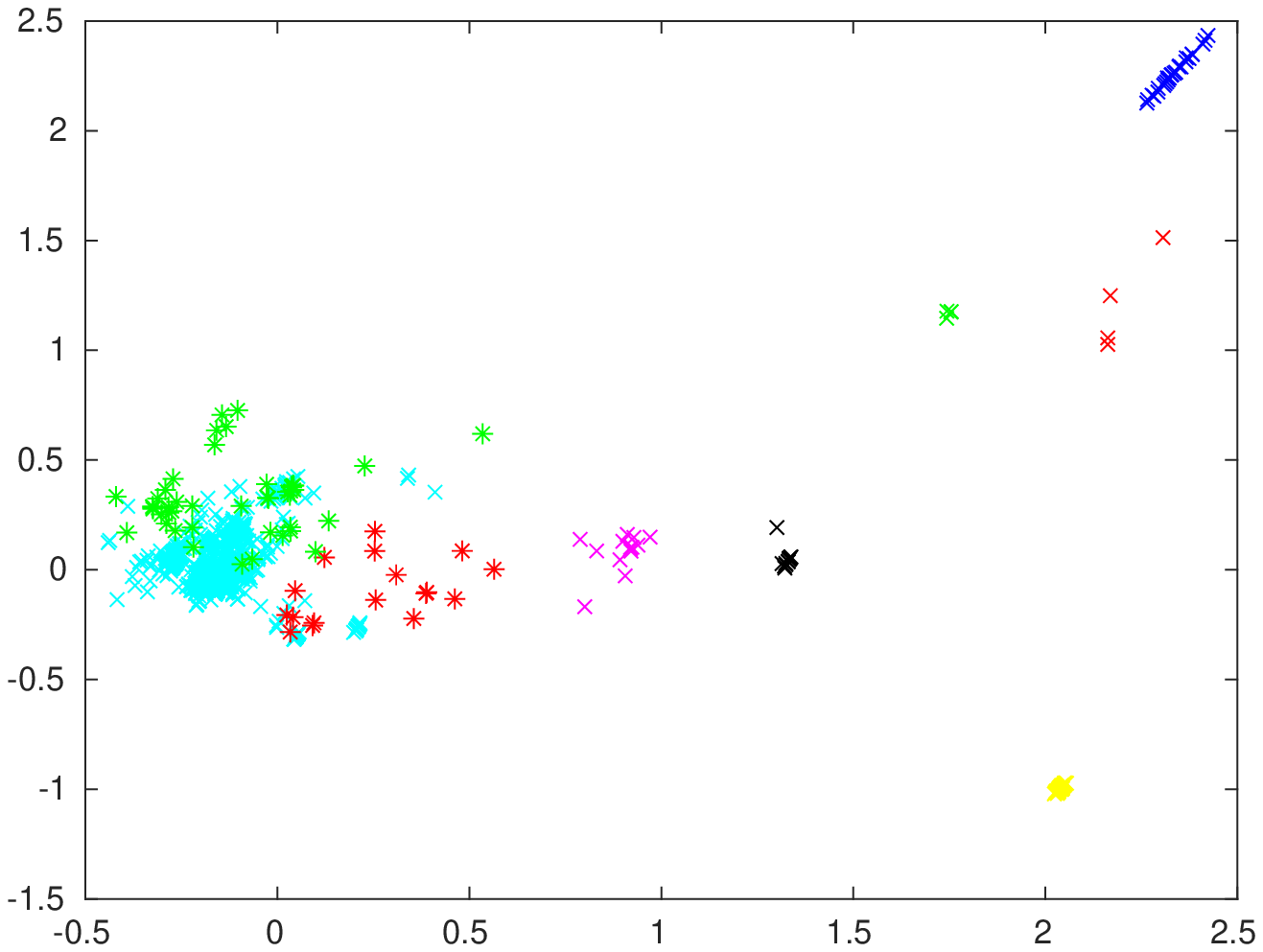} }}
    \subfloat[Iteration 53]{
    		\label{Fig:FinalClustering}
	    \includegraphics[scale=0.3]{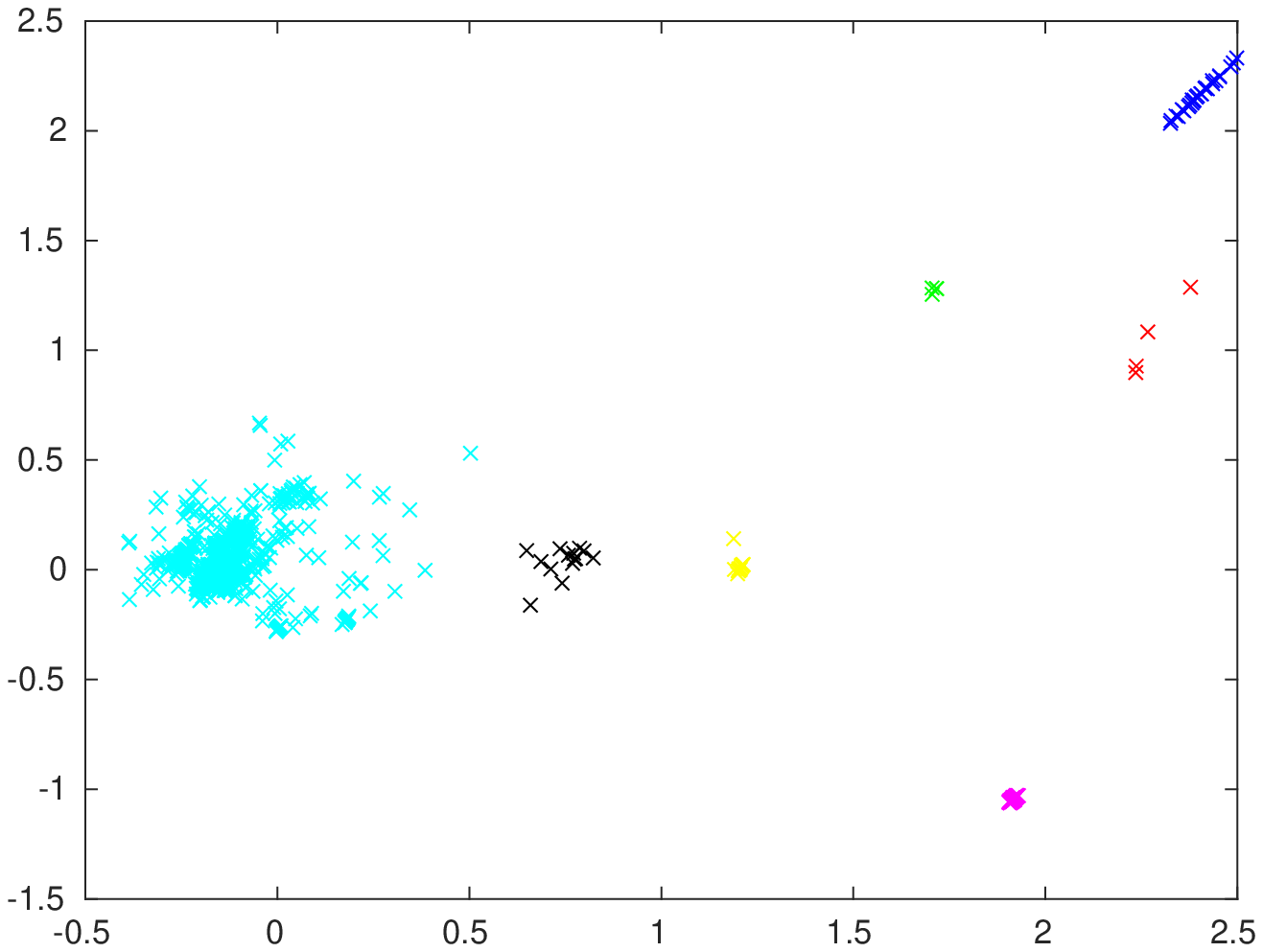} 
    }
    \caption{Clusterings generated by $imwk$-means on the data sets produced by our method, at each iteration. These figures show the separation of clusters until the convergence of our method (iteration 53). In the last iteration the average silhouette index is of $0.92$.}%
	\label{Fig:NewMethodIterations}	
\end{figure*}

All of the above is very positive, but we need to ensure that our final clustering (ie. Figure \ref{Fig:FinalClustering}) is actually meaningful in the real-world. In order to do this, we analysed each cluster with the help of VirusTotal \cite{VirusTotal}. The latter is the most prominent online public service with multiple anti-virus scanners \cite{sakib2020maximizing}. Even with its help, analysing our malware data is far from being a trivial task. VirusTotal 
is capable of describing each malware in our data set by employing the use of a number of AntiVirus (AV) software. However, this does not mean that each and every AV will agree what a malware sample actually is (or even if the sample is really a malware). Also, different AVs may have different taxonomies. Thus, even if two AVs agree what a malware sample actually is, they may use different names. 

Thanks to our analysis we can describe each cluster as follows:

Cluster one: this cluster contains four malware samples, Figure \ref{Fig:FinalClustering} shows these in red. The AV labels for these in VirusTotal seem to be quite different (using names like adware and trojan, which are not mutually exclusive), however, all four samples have exactly the same compilation timestamp. This certainly suggests these malware samples are very much related.

Cluster two: this cluster contains four malware samples, Figure \ref{Fig:FinalClustering} represents these in green. The vast majority of AV's in VirusTotal labels these as trickbot, a trojan designed to steal banking information in particular. All the malware samples in this cluster share the same import hash (imphash), which means they have very similar import tables and are by consequence similar.

Cluster three: this cluster contains 27 malware samples, Figure \ref{Fig:FinalClustering} represents these in blue. These malware samples can be easily characterised by the huge amount of behaviour they exhibit while running in a Virtual Machine, sometimes spawning over 100 processes. According to VirusTotal, over 30 AVs (out of 60) label most of the samples in this cluster as generic malware. Around five AVs consistently label these malware samples with ransomware characteristics, while other five AVs use the term flystudio (adware).

Cluster four: this cluster contains 14 malware samples, Figure \ref{Fig:FinalClustering} represents these in black. Eight of the malware samples appear to be a specific type of trojan (a downloader/installer). All but one of the malware samples has the same compilation timestamp, and share the same imphash.

Cluster five: this cluster contains 14 malware samples, Figure \ref{Fig:FinalClustering} represents these in yellow. These malware samples have different imphashes but the AVs in VirusTotal labels all of these samples as belonging to the ransomware family GandCrab. 

Cluster six: this cluster contains 79 malware samples, Figure \ref{Fig:FinalClustering} represents these in magenta. Almost all malware samples in this cluster exhibit extremely similar behaviour given they all share the same imphash value. VirusTotal suggests these as adware in general, and AVs classify them as being in the Adposhel or DNSUnlocker families. 

Cluster seven: this cluster contains 1,858 malware samples, Figure \ref{Fig:FinalClustering} represents these in cyan. This is a very large cluster for us to analyse each and every malware sample in VirusTotal. The malware samples in this cluster seem to be associated with different families, but also seem different from the malware samples in other clusters. 

The above shows there is a considerable difference in cardinality between clusters, which is certainly expected. Malwares have a tendency to appear in bursts, and their distribution is highly skewed \cite{song2016learning}. Our method will identify the most anomalous clusters first. Further analysis could be done by applying our method solely to the data in the largest cluster. We do not pursue this here because we have already clearly achieved our aim. Taking the cluster's descriptions above together with: (i) the average silhouette of $0.92$ given by $imwk$-means; (ii) the increased average silhouette of $k$-means++ (from $0.39$ to $0.52$, the latter on the data set generated by icdf); (iii) the mathematical model shown in Section \ref{Sec:NewMethod}, we can state our method produces a data set which increases the chances of a meaningful clustering. 


\section{Conclusion}
\label{Sec:Conclusion}

In this paper we faced the problem of finding meaningful clusters in drive-by-download malware data. The patterns in this type of data can be difficult to identify, particularly if using a distance based clustering algorithm (see Figures \ref{Fig:OriginalData} and \ref{Fig:kmeans}). We identified as the main reason for this the fact that classical data normalisation treats all features equally, instead of favouring those that are more relevant.

In order to address the above, we introduced a data pre-processing method called iterative cluster-dependent feature rescaling (for details see Section \ref{Sec:NewMethod}). This method makes use of cluster-dependent feature weights to iteratively separate the clusters in a data set (see Figure \ref{Fig:NewMethodIterations}). This mathematically sound method leads to higher average silhouettes. For instance, $k$-means++ saw an increase from $0.39$ to $0.52$ when using our feature rescaling method, while $imwk$-means went as high as $0.92$. Hence, more meaningful clusters.

We foresee our method being used in the data pre-processing stage of a malware clustering task, or perhaps even in other clustering tasks. In the future we intend to investigate the use of this method in supervised classification problems.

%
%
\bibliographystyle{elsarticle-num}
\bibliography{references.bib}
\end{document}